\documentclass[aps,prl,twocolumn,superscriptaddress,showpacs,showkeys]{revtex4}
\usepackage{graphicx}
\usepackage{dcolumn}
\usepackage{bm}

\begin{document}

\preprint{APS/123-QED}

\title{Ultra-high magnetic field study of the layer split bands in Graphite}

\author{R.J. Nicholas}
\email{r.nicholas1@physics.ox.ac.uk}
\affiliation{Dept. of
Physics, University of Oxford, Clarendon Laboratory, Parks Rd.,
Oxford, OX1 3PU, U.K.}

\author{P.Y. Solane}
\affiliation{Laboratoire National des Champs Magnetiques Intenses,
CNRS-UJF-UPS-INSA, 143 avenue de Rangueil, 31400 Toulouse, France}

\author{O. Portugall}
\affiliation{Laboratoire National des Champs Magnetiques Intenses,
CNRS-UJF-UPS-INSA, 143 avenue de Rangueil, 31400 Toulouse, France}

\date{\today}
\begin{abstract}
We report studies of the magnetospectroscopy of graphite into a new regime of high energies and ultra-high magnetic fields which allows us to perform the first spectroscopic studies of the interlayer split off bands, $E_{1}$ and $E_{2}$. These bands can be well described by an asymmetric bilayer model and have only a small interlayer band gap asymmetry. We show that all of the properties of the electrons and holes can be described by a simple relativistic behaviour determined by $\gamma_{0}$ and $\gamma_{1}$.
\end{abstract}

\maketitle

The recent surge in interest in two dimensional electronic systems
formed from monolayer \cite{novos04, novos05, zhang:2005,
zhang:2006} and bilayer graphene\cite{novosolev:2006,
mccann:2006b, guinea:2006}, is based on the properties of bulk
graphite\cite{orlita:2008, orlita:2009, koshino:2008, zhou:2006,
gruneis2008a,gruneis2008b}. In particular the many exciting properties of
bilayer graphene are crucially dependent on understanding the
interlayer coupling that originates in bulk graphite. There is mounting evidence\cite{partoens:2007,koshino:2008,koshino:2009,orlita:2009} that the
majority of the properties of graphite can be described quite
simply at the high symmetry points of the Brillouin zone by a
combination of a single layer graphene (SLG) model for massless Dirac fermions at the $H$-point
and a bilayer (BLG) model for massive particles at
the $K$-point. Spectroscopic and theoretical estimates of the band parameters still remain controversial however, particularly around the $K$-point, as only very limited experimental evidence exists of the properties of the interlayer split off bands, $E_{1}$ and $E_{2}$ formed by the interlayer coupling which is dominated by the matrix element $\gamma_{1}$ in Bernal stacked layers. By using ultra-high magnetic fields we now extend the magnetospectroscopy of graphene and graphite into a new regime of high energies which allows us to perform the first magnetospectroscopy of the interlayer split off bands and show that their behaviour can be modelled very well by the analogue of relativistic behaviour predicted by the BLG model.

\begin{figure}[tbp]
\begin{center}
\includegraphics[width=0.95\linewidth]{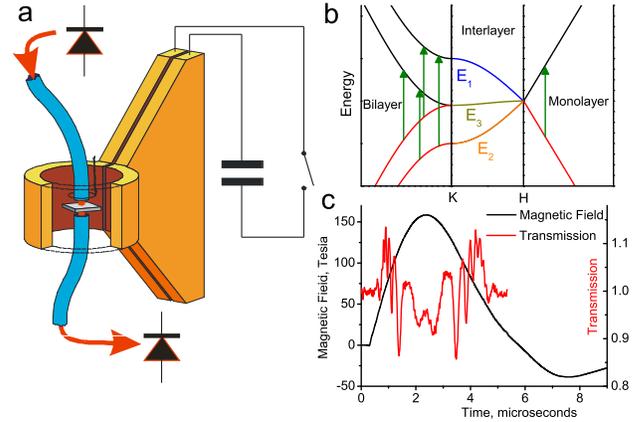}

\caption{\label{schema} (Color online) a) Schematic view of the experimental system, b) graphite band structure c) Typical time dependence of the magnetic field and sample transmission for 1220nm radiation.}

\end{center}
\end{figure}

Traditionally the band structure of graphite has been described by
the Slonczewski-Weiss-McClure (SWM) tight binding
model\cite{slonczewski:1958, mcclure:1960} which requires the use
of seven tight binding parameters $\gamma_{0}$....$\gamma_{5},
\Delta$ determined by interlayer and intralayer matrix
elements. This provides a description of the dispersion relations
all around the Brillouin zone edge from the hole pocket centered
at the $H$-point to the electrons around the $K$-point. By contrast, the SLG/BLG model uses $\gamma_{0}$ with only a single interlayer matrix element, $\gamma_{1}$, and is found to provide a remarkably accurate \cite{orlita:2009} description of the
magneto-optical properties of graphite by assuming that this is dominated by the $H$- and $K$-points. Even at low energies where the band structure is known to be more complex\cite{gruneis2008a, schneider:2009}, the slight
modification of introducing an asymmetric velocity for the
$K$-point fits the data very well\cite{chuang:2009} and can describe most of the behaviour predicted in the SWM
model. In the simplified bilayer picture the band structure is shown in Fig.\ref{schema}b, with the K-point having two touching symmetric massive bands and split off bands at $\pm2\gamma_{1}$. Introduction of the asymmetric bilayer model (ABM) (as occurs due to the presence of $\gamma_{4}$ in the SWM) means that $E_{2}$-$E_{3}^{+}$ and $E_{1}$-$E_{3}^{-}$ become symmetric pairs of bands each with their own respective Fermi velocity, $v_{F}^{+}$ and $v_{F}^{-}$ which are predicted to differ by $\sim$10\% in both the SWM model and density functional theory.

\begin{figure*}
\begin{center}
\includegraphics[width = 17cm]{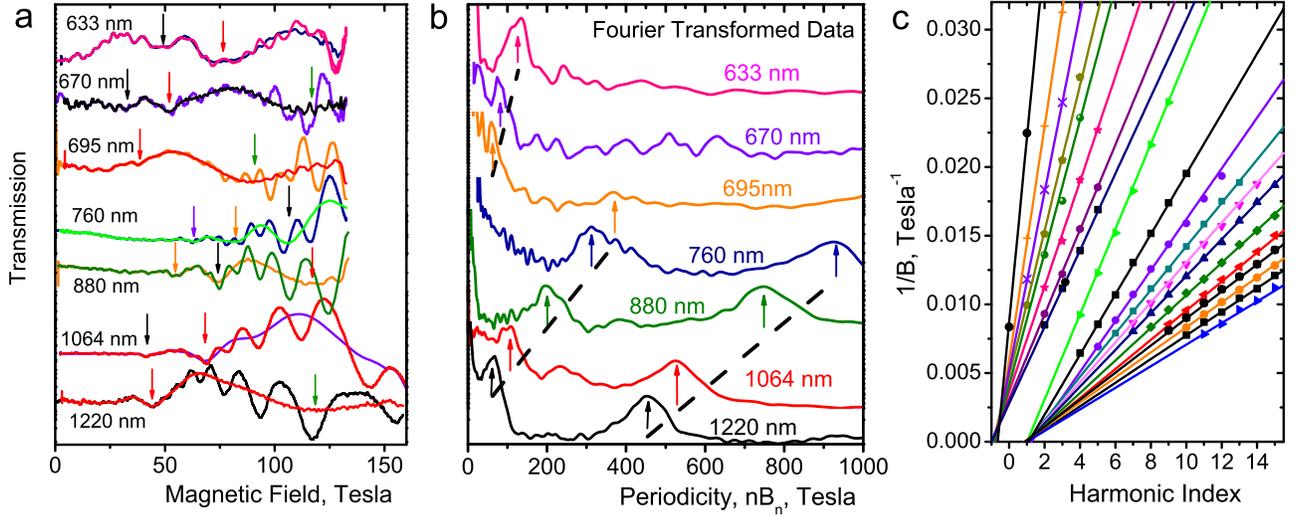}
\end{center}
\caption{(Color online) a) Magnetotransmission as a function of magnetic field (original data and spectra filtered to remove $E_{3}\pm$ transitions) for a series of wavelengths,  b) Fourier Transforms as a function of 1/B for the traces shown in a c) Plots of the values of 1/$B_{n}$ versus (arbitrary) harmonic index for different transitions.}
\label{absorptionFT}

\end{figure*}

Measurements were made of the transmission of thin ($\sim$20nm)
exfoliated natural crystalline graphite up to magnetic fields of 160 T at 300K. Fields were generated using a semi-destructive single-turn coil technique shown schematically in Fig. \ref{schema}a that provides pulse lengths of 5 $\mu$s. The transmission of a series of lasers in the region 1550 - 630 nm
was detected using high speed photodiodes and a 200 MHz low
noise amplifier to follow complete cycles of the pulsed magnetic field, allowing the rising and falling parts of the magnetic field cycle to be compared and averaged. Typical recordings of the time dependent magnetic field and transmission are shown in Fig. \ref{schema}c.

Typical experimental results are shown in Fig.
\ref{absorptionFT} where there are a series of absorptions which we will show are periodic in 1/B with more than one periodicity, depending on the photon energy.  The strongest
high frequency series are assigned to the ($\Delta n=\pm1$) transitions between the $E_{3}^{\pm}$ $K$-point Landau levels, by extrapolation from
previous work \cite{orlita:2009, ubrig:2011}.  The lower frequency
resonances are, however, previously un-reported, and are due to transitions from and to the split-off bands.

We analyse the magneto-optical response by introducing a
Fourier analysis procedure which allows us to isolate the in-plane
quantization effects and separate out the contributions from the different bands when working at fixed photon energy. The periodicity of the transitions in reciprocal magnetic field can be seen from an analysis starting from the bilayer approach for the $K$-point\cite{koshino:2008, mucha2011}, where we write

\begin{eqnarray}
  \hspace{-5mm}\varepsilon_{n} =&&
  \frac{s}{\sqrt{2}}\bigg[(\lambda\gamma_{1})^{2}+(2n+1){\Delta_{B}}^{2}+\nonumber\\
&& \hspace{-5mm}
\mu\sqrt{(\lambda\gamma_{1})^{4}+2(2n+1)(\lambda\gamma_{1})^{2}{\Delta_{B}}^{2}+{\Delta_{B}}^{4}}\,\bigg]^{1/2},
  \label{E1}
  \end{eqnarray}

\noindent where $\Delta_{B}$ is the magnetic energy for the
graphene-like in-plane motion

  \begin{equation}
  \Delta_{B}= v^{\pm}\sqrt{2\hbar e B} = \sqrt{\alpha B},
  \label{E2}
  \end{equation}

\noindent where $v^{\pm} = \sqrt{3} e a_{0} \gamma_{0}/2\hbar$ (with different values for the electrons and holes in the ABM), s =
$\pm$ for the electrons and holes, and $\mu$ = $\mp$
corresponds to the fundamental and split-off
bands. For the $K$-point transitions in graphite $\lambda$=2.
The Landau level energies are given by

\begin{eqnarray}
  \hspace{1mm}(n^{2}+n)\alpha B_{n} =(n+1/2)\varepsilon_{n}^{2}\pm\nonumber\\
&\hspace{-40mm}
\lambda\gamma_{1}\varepsilon_{n}(n^{2}+n)^{1/2}\sqrt{(1+\frac{\varepsilon_{n}^{2}}{4(n^{2}+n)(\lambda\gamma_{1})^{2}})},
  \label{E3}
  \end{eqnarray}

For the dominant interband optical transitions with a selection rule
$\delta n$ = $\pm1$ the transition energies are

\begin{eqnarray}
  \hspace{-5mm}E = |\varepsilon_{n}^{+}| + |\varepsilon_{n\pm1}^{-}|
  \label{E4}
  \end{eqnarray}

Working at constant energy E for high Landau level indices $n$
Eq. \ref{E3} and Eq. \ref{E4} approximate to give the magnetic field dependence of transitions
between the nth and (n-1)th levels $B_{n}$ as the
remarkably simple expression

\begin{eqnarray}
  \hspace{-5mm}n\alpha B_{n}=&&
  E^{2}/4  +  2\gamma_{1}E/2,
  \label{E5}
  \end{eqnarray}

where $\alpha$ is chosen as the appropriate mean of the values deduced from $v$ for the bands involved in the transitions.

\begin{figure}[tbp]
\begin{center}
\includegraphics[width=0.95\linewidth]{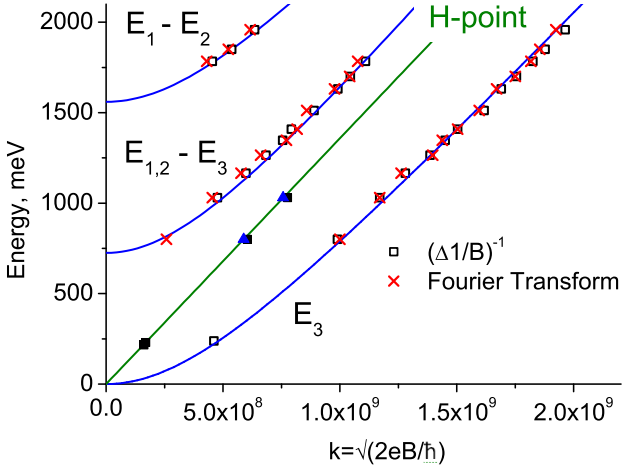}

\caption{\label{periodicity} (Color online)Energy momentum relation for interband excitations in graphite deduced from the periodicity of the magneto-oscillations and from Eq. \ref{E8}, low energy (229meV) data from \cite{plochocka2012}. The H-point dispersion comes from n=1 to n=2 interband transitions giving v=1.03 $\times$10$^{6}$m/s\cite{sadowski:2006}.}

\end{center}
\end{figure}

 The values for $B_{n}$ predicted by equation \ref{E5} are within $<0.5\%$ of the values predicted by the full bilayer model
for $n>3$, using the same values for $\gamma_{0}$ and
$\gamma_{1}$, suggesting that the observed behaviour should be
very accurately periodic in 1/B and can be used to determine these
parameters without the need for qualitative judgements of the fitting
accuracy for multiple peaks. 

For transitions involving the layer split-off
 bands we have a similar expression with similar accuracy.

\begin{eqnarray}
  \hspace{-5mm}n\alpha B_{n}=&&
  (E-\Gamma)^{2}/4  +  2\gamma_{1}(E - \Gamma)/2,
  \label{E7}
  \end{eqnarray}

where $\Gamma$ = $2\gamma_{1}$ for transitions between the layer split-off bands and $E_{3}$ ($E_{2} \rightarrow E_{3+}$, $E_{3-} \rightarrow E_{1}$) and $\Gamma$ = $4\gamma_{1}$ for transitions directly between the two layer split-off bands ($E_{2} \rightarrow E_{1}$).

These simple expressions are a consequence of the
intermediate energy zero field result \cite{mccann:2006b} for a
bilayer:

\begin{eqnarray}
  \hspace{-5mm}v^{2}p^{2}=&&
  \varepsilon^{2}  +  2\gamma_{1}\varepsilon,
  \label{E6}
  \end{eqnarray}

where the momentum $p^{2}$ has been quantized at $2ne\hbar B$ and
the transition energy E=2$\varepsilon$.  This
expression is an exact analogy of the relativistic energy momentum
relationship where $\gamma_{1}$ plays the role of the particle
rest mass\cite{tung2011}. Including the split off bands gives the set of relativistic dispersion relations:

\begin{eqnarray}
  \hspace{-1mm}\varepsilon_{3}^{\pm}=\pm\sqrt{\gamma_1^2 + p^{2}v^{2}_{\pm}} \quad\mp\,\gamma_1\nonumber\\
    \hspace{-1mm}\varepsilon_{1,2}^{\pm}=\pm\sqrt{\gamma_1^2 + p^{2}v^{2}_{\mp}} \quad+\,\delta
 \label{E8}
  \end{eqnarray}

The pairs of bands are asymmetric which is a feature of both the SWM and ab initio calculations using density functional theory\cite{gruneis2008b} which predict $v_{+}/v_{-}\approx 1.12$. Both types of theory also predict that the $E_{2}-E_{3}$ and $E_{3}-E_{1}$ interlayer gaps should be asymmetric. The SWM model predicts for example that ($E_{2}+E_{1}-2E_{3}$) = 2($\Delta -2\gamma_{2} + 2\gamma_{5}$) = 2$\delta$, with typical fitting parameter values from the literature suggesting a wide range of values for 2$\delta$ in the region 0.06 to 0.3 eV \cite{gruneis2008a,gruneis2008b,nakao:1979,tung2011} and tight binding models predicting 0.1-0.2 eV\cite{gruneis2008b}. All of the transition energies plotted in Fig. \ref{periodicity} can then be calculated from the differences of the energies in Eq.\ref{E8}.

\begin{figure}[th!]
\begin{center}
\includegraphics[width=0.95\linewidth]{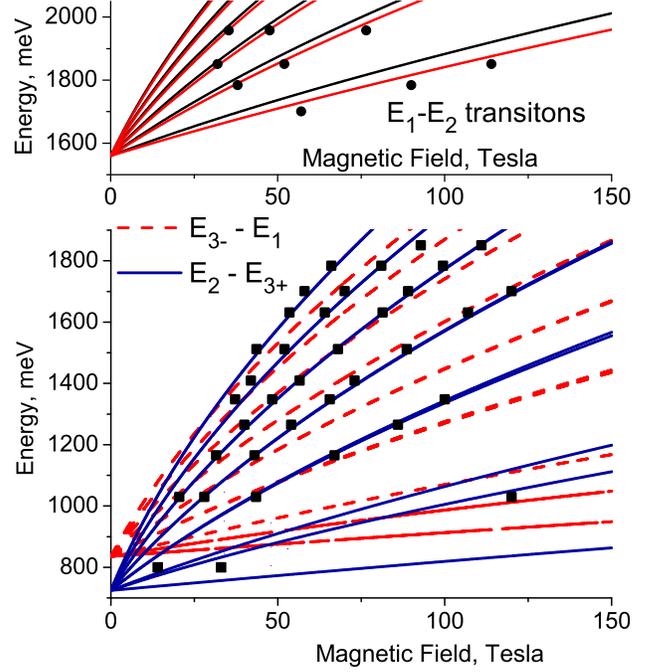}
\caption{\label{E1E3}(Color online)Transition energies as determined from the Fourier filtered plots shown in Fig.\ref{absorptionFT} and calculated from the ABM.}
\end{center}
\end{figure}

For a fixed photon energy the magnetic field values for the transitions are thus
expected to show a very well defined periodic dependence on 1/B
with a periodicity ($n\alpha B_{n}$=1/$\Delta (1/B_{n})$) given by Equations \ref{E5}, \ref{E7}.  The experimental results demonstrate that this prediction holds extremely well. Plots of 1/B values for the absorption minima show (Fig. \ref{absorptionFT}c) a very accurate harmonic series and Fourier transforms of the traces in Fig. \ref{absorptionFT}a as a function of 1/B (Fig \ref{absorptionFT}b) show well defined peaks.  The highest periodicity peak (300-1300T) corresponds to the the well known $K$-point $E_{3}^{\pm}$
transitions.  The transforms show the appearance of two new peaks for energies more than 1 eV (1220 nm) (100-400T) and 1.8 eV (690 nm) (50-200T) respectively. The new series correspond to transitions involving the
split off bands, the first from a combination of $E_{2}-E_{3}^{+}$ and $E_{3}^{-}-E_{1}$, and the second directly from $E_{2}$-$E_{1}$.  The periodicities of all three transitions are used to give the k-vector (=$\sqrt{2enB_{n}/\hbar}$) and then plotted as energy-k-vector dispersions in fig.\ref{periodicity} with the periodicities determined both
from 1/B plots and from the Fourier transforms. The $E_{3}^{-} - E_{3}^{+}$ transitions show excellent agreement with Eq. \ref{E5} using typical values for $\gamma_{0}$=3.18 eV ($v$=1.03 $\times$10$^{6}$m/s) and $\gamma_{1}$=0.39 eV\cite{sadowski:2006, orlita:2009}. Since these transitions correspond to high quantum numbers, $n$, they measure the average of the values for $E_{3+}$ and $E_{3-}$ (which we later show to be: $v^{+}$=1.14 $\times$10$^{6}$m/s and $v_-$=0.92 $\times$10$^{6}$m/s) and it is impossible to resolve any electron-hole splitting. Fitting the periodicities for the $E_{2}$-$E_{1}$ transitions with Eq.\ref{E7} also gives good agreement with the same parameter values for the average $v$ and a total gap of 4$\gamma_{0}$. The results for the $E_{3}-E_{1,2}$ transitions are more surprising, however, as these require a significantly higher value for $v$ and a reduced value for the band gap of 0.725 eV. This suggests that transitions from $E_{2}-E_{3+}$ are dominant, where both bands have the higher Fermi velocity and the band gap is reduced due to the asymmetric interlayer coupling\cite{gruneis2008b},\cite{slonczewski:1958},\cite{mcclure:1960}.

The existence of a well defined periodicity in 1/B allows us to examine the split-off band transitions in more detail by using a Fourier blocking filter in 1/B to remove the higher frequency oscillations from the (E$_{3}^{\pm}$) transitions, as shown in Figure \ref{absorptionFT}. For each recording a second trace is shown where the data has been processed with a Fourier blocking filter over the range ($nB_{n}(E_{3}^{\pm})\pm200T$) which removes the oscillations associated with the $E_{3}^{\pm}$ transitions. We first analyse the $E_{2}-E_{1}$ transitions which are compared directly with the exact ABM predictions for low quantum numbers as shown in Fig. \ref{E1E3}. This allows us to make an accurate fit for $\gamma_{1}$=0.39 eV, since this is the only parameter which enters into the $E_{2}-E_{1}$ separation, even in the full SWM model. 

As already suggested above, the second series of transitions from $E_{3} \leftrightarrow E_{1,2}$ require an increased value of $\gamma_{0}$=3.63 eV ($v^+$=1.14 $\times$10$^{6}$m/s) when fitted with either Eq.\ref{E7} or the full asymmetric bilayer model Eq.\ref{E3}, with a suitably reduced value of $v_-$.
This suggests that at high fields the transitions
are dominated by the symmetric pair of bands $E_{2}$ and $E_{3}^{+}$, which have a higher electron velocity due to the band asymmetry associated with the $\gamma_{4}$ term and the free electron contribution \cite{plochocka2012} in the SWM model. The two velocities are slightly more asymmetric than observed previously in low field measurements on graphite\cite{chuang:2009} and for monolayer\cite{Deacon:2007} and bilayer \cite{henriksen:2008} graphene, which is not unexpected given that our measurements use a larger range of energies. Another potential explanation is that Coulomb many-body interactions are becoming larger at higher energies \cite{shizuya2011a}. Fig.\ref{E1E3} illustrates the exact transition energies calculated using the asymmetric bilayer model for all of the layer split band transitions with a value for ($E_{2}+E_{1}-2E_{3}$) = 0.11 $\pm$ 0.04 eV. There are very few direct measurements of the asymmetry but a value of 0.13 eV was observed by Bellodi et al\cite{bellodi1975} from thermoreflectivity measurements at zero field.  These measurements study transitions which are strongly influenced by the Fermi level occupancy of the states around the $K$-point which makes their interpretation difficult\cite{dresselhaus:1976} but they are nevertheless in good agreement with our measurements, which suggests that the layer split band gaps $E_{1}-E_{3}$ and $E_{3}-E_{2}$ are significantly less asymmetric than often thought and should provide a good constraint for values used in fitting theoretical models of graphite.  Our value for $E_{3}-E_{2}$ of 0.725$\pm0.02 eV$ is also in good agreement with that reported from ARPES measurements (0.71$\pm0.015 eV$)\cite{gruneis2008a}, although this method could not observe the upper band. Split off band transitions have been observed with bilayer graphene where electroreflectance measurements\cite{kuzmenko:2009} measure split off bands at half the energy for graphite due to the single sided coupling. The resonances at 0.363 and 0.393 eV are slightly less asymmetric than our graphite values where considerably more interlayer coupling is occurring.  

Perhaps the most remarkable conclusion is that the dispersion relation for all of the interband excitations can be deduced from the simple quasi-relativistic dispersion relations Eq.\ref{E8} with slightly more asymmetric velocities than reported previously\cite{chuang:2009},\cite{Deacon:2007},\cite{henriksen:2008}. In the visible region of the spectrum all of the excitations can be considered to be in the relativistic region of the dispersion relation. 

\section{acknowledgments} Part of this work has been supported by
EuroMagNETII under the EU contract FP7-INFRASTRUCTURES-2008-228043 of the
7th Framework 'Research Infrastructures Action'.



******************
\end{document}